\documentclass[journal]{IEEEtran}
\usepackage{tikz}
\usepackage{pgfplots}
\usepackage{pgfplotstable}
\usepgfplotslibrary{fillbetween, patchplots, polar}
\tikzset{>=latex}
\usetikzlibrary{
  3d,
  arrows,
  arrows.meta,
  calc,
  matrix,
  plotmarks,
  positioning,
  shadings,
  shapes.geometric,
  decorations.pathmorphing
}
\tikzset{
  currentarrow/.style={
    -{Stealth[length=2.2mm,width=0.8mm]},
    line width=0.6pt,
    line cap=round
  }
}

\pgfplotsset{compat=newest}
\usepackage{standalone}    %
\newlength\figurewidth
\newlength\figureheight
\usepackage{amsmath,amssymb,amsfonts}
\usepackage{graphicx} 
\usepackage{cite}
\usepackage{xcolor}
\newcommand{\tM}[1]{\widetilde{\mathbf{#1}}}

\providecommand*{\M}[1]
{\mathbf#1}        
\providecommand*{\V}[1]{\boldsymbol#1} 
\providecommand*{\UV}[1]{\hat{\boldsymbol#1}}  

\providecommand*{\T}[1]{\mathrm{#1}}  
\providecommand*{\ju}{\ensuremath{\T{j}}} 
\providecommand*{\diff}{\operatorname{d}\!}  

\newcommand{\regT}{\varOmega_\T{s}}
\newcommand{\regR}{\varOmega_\T{o}}

\begin{document}

\title{
Degrees of Freedom Sampling
}

\author{Jiawang Li, Mats Gustafsson

\thanks{This work was supported in part by a Project Grant in ELLIIT 
Call  D, in  part  by  NextG2Com  (grant  no.  2023-00541) 
funded by the VINNOVA program for Advanced Digitalisation, and in part by Swedish Research Council SEE-6GIA 2024-06482.  (Corresponding author: \textit{Jiawang Li}).}%
\thanks{Jiawang Li and Mats Gustafsson are with the Department of Electrical and Information Technology, Lund University, 22100 Lund, Sweden (e-mail: {\{jiawang.li, mats.gustafsson\}}@eit.lth.se).
}%

}

\maketitle

\begin{abstract}
A geometry-based framework for sampling electromagnetic fields over surfaces is presented. The approach extends the shadow-area formulation of electromagnetic degrees of freedom (DoF) from a global mode count to a spatially resolved DoF density that determines the sampling distribution. The observation surface is partitioned into cells containing approximately one DoF, with the cell area determined by the DoF density and the cell shape determined by the projection of the DoF density vector onto the observation surface. In the far field, the source shadow area defines a directional DoF density over the observation
sphere, while in the near field, the mutual-shadow density and a DoF density direction determine the local sampling geometry. The resulting method requires only the source–observation geometry and does not require an explicit singular-value decomposition of the propagation operator. Numerical results for several source and observation
geometries demonstrate reconstruction performance close to that of operator-based sampling methods. A beamforming interpretation further shows that the sampling cells correspond to geometry-dependent beam regions, establishing a unified connection between DoF, sampling, and beamforming.
\end{abstract}

\begin{IEEEkeywords}
Degrees of freedom, sampling, mutual-shadow density, near-field
\end{IEEEkeywords}

\section{Introduction}
\IEEEPARstart{M}{easuring} an electromagnetic field over a two-dimensional
(2D) surface requires representing a continuous field with
a finite set of spatial samples. In antenna characterization, electromagnetic imaging, and inverse source reconstruction, dense regular grids remain common because they are simple to implement and compatible with conventional processing methods~\cite{mezieres2021antenna,fuchs2017fast,
capozzoli2016singular,randazzo2021two,maisto2021near,marengo2000inverse,
han2026reactive,solimene2018inverse,bucci1997electromagnetic}. A common choice is a spacing of approximately one half wavelength~\cite{yaghjian1986overview,
marks1993advanced}. However, such uniform sampling does not account for the finite source extent or its geometry relative to the observation region and may therefore introduce substantial redundancy when only a limited number of independent spatial modes can be supported~\cite{miller2000communicating,gabor1961iv,bucci1989degrees}.

This motivates formulating the sampling problem in terms of the effective dimension of the electromagnetic field space. The electromagnetic degrees of freedom (DoF) quantify the number of independent spatial modes that can couple a source region to an observation region~\cite{franceschetti2017wave,Bucci2025,
miller2000communicating,poon2005degrees,dardari2020communicating,puggelli2025maximizing,yuan2024breaking,li2026degrees,del2026effective}. Existing characterizations include Weyl-type asymptotic estimates~\cite{weyl1911asymptotische}, nonredundant field representations~\cite{bucci1989degrees,bucci1998representation, maisto2021efficient,maisto2021near,pizzo2022nyquist,kunsch2005optimal, agrell2017multidimensional}, singular-value analysis of propagation operators~\cite{miller2000communicating,poon2005degrees}, and paraxial approximations~\cite{miller2019waves,piestun2000electromagnetic}. More recently, the mutual-shadow framework has related both the DoF and the spectral transition of electrically large propagation systems directly to geometric quantities~\cite{gustafsson2025shadow,gustafsson2025degrees}. These formulations provide a global DoF count but do not describe how
the available spatial information is distributed over an extended observation region.

Nonredundant sampling methods reduce measurements by exploiting the geometry-dependent spatial bandwidth of the field rather than imposing a fixed physical spacing~\cite{bucci1989degrees,bucci1998representation}. Geometry-adapted coordinates and warping transformations have been used to construct compact sampling grids on planar, cylindrical, and more general
surfaces~\cite{maisto2021efficient,maisto2021near}. Related approaches derive reduced discretizations from the asymptotic or spectral properties of the radiation operator~\cite{pierri2020asymptotic,pierri2021ndf,
leone2022dimension,solimene2019sampling,migliore2025intuitive}, while compressed sensing can provide additional reduction when sparsity is available ~\cite{candes2008introduction,bangun2022optimizing}. These
approaches, however, generally rely on specific parameterizations, geometries, or prior spectral information.

In this paper, we develop a geometry-based framework for far- and near-field electromagnetic sampling using the mutual-shadow DoF measure~\cite{gustafsson2025shadow,gustafsson2025degrees}. Rather than using the spatial DoF only as a global estimate of the number of required modes, the proposed framework exploits its spatial or directional distribution to determine where the field should be sampled. To evaluate the resulting sampling distributions, we consider both spectral- and Frobenius-norm reconstruction errors. The spectral norm characterizes the worst-case reconstruction error, whereas the normalized Frobenius norm reflects the average reconstruction performance over the spatial modes. The corresponding singular-value bounds provide references for the best achievable accuracy under the two criteria.

For far-field observations, the sampling problem involves determining not only the required number of angular samples but also their distribution over the observation sphere to enable efficient reconstruction of the directional field. In this regime, the projected shadow area of the source defines a directional DoF density over the sphere. Sampling points are then constructed by partitioning the sphere into cells with approximately equal mutual-shadow contribution, such that each cell represents approximately the same fraction of the spatial DoF. The resulting sampling distribution is further evaluated from a beamforming perspective, where the sampling directions are used as beam-steering directions to examine the corresponding beam coverage and spatial selectivity over the observation sphere.

For near-field observations, the mutual-shadow formulation defines a spatial DoF density over the observation surface. The density determines the required sampling-cell area, while a DoF density direction obtained from the complete source aperture determines the cell orientation and aspect ratio. The resulting DoF sampling construction therefore adapts both the density and anisotropy of the sampling pattern to the source--observation geometry. Numerical singular-value, reconstruction-error, and beamforming results are used to verify that the geometry-derived sampling distributions retain the dominant spatial information of the continuous electromagnetic field.

The rest of this paper is organized as follows. Section~\ref{sec:Modelling and error formulation} formulates the sampling problem and introduces the reconstruction errors. Section~\ref{sec:far_field_sampling} develops the mutual-shadow-based far-field sampling formulation. Section~\ref{sec:Near-field surface sampling} extends the framework to near-field surface sampling. Finally, Section~\ref{sec:Conclusions} concludes the paper.

\textit{Notation:} Throughout this paper, boldface letters indicate vectors and boldface uppercase letters designate matrices. Superscript $(\cdot)^{\T{H}}$ and $(\cdot)^{+}$ stand for Hermitian transpose and pseudo-inverse~\cite{ben2003generalized}.

\section{Modelling and error formulation
}\label{sec:Modelling and error formulation}

We consider an electromagnetic field generated by a continuous source distribution on a convex surface and sampled over an observation surface. The objective is to reconstruct the field from a finite set of samples while retaining the relevant information of the source–observation interaction. For clarity of exposition, the analysis is presented using a scalar-field model~\cite{miller2000communicating,maisto2021efficient,leone2022dimension,Bucci2025}, while the proposed sampling methodology can be readily extended to vector electromagnetic fields. 
\begin{figure}[t]
  \centering
\includegraphics[width=1\linewidth]{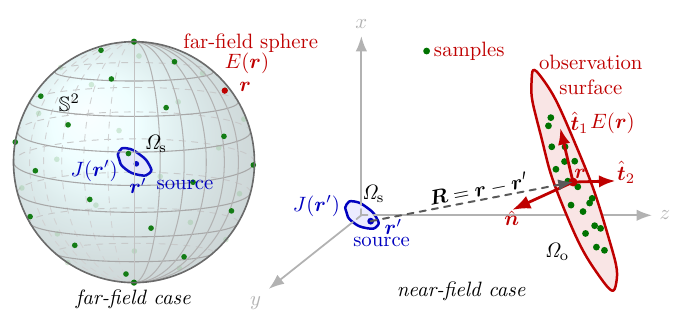}
\caption{Geometry of the electromagnetic sampling problem. In the far field (left), the observation region reduces to the unit sphere, $\regR=\mathbb{S}^2$, whereas in the near field (right), $\regR$ denotes a general observation surface.}
\label{fig:research_problem_surface}
\end{figure}
The considered geometry is illustrated in Fig.~\ref{fig:research_problem_surface}. 
The source current $J(\V r^\prime)$ is defined over the source surface $\regT$, and the resulting field $E(\V r)$ is observed and sampled over the observation domain. For a general observation surface $\regR$, a source point $\V r^\prime$ and an observation point $\V r$ are connected by the displacement vector $\V R=\V r-\V r^\prime$, while the local surface orientation is characterized by the tangential directions $\UV t_1$ and $\UV t_2$ and the unit normal $\UV n$. In the far-field specialization, the observation domain is represented by the unit sphere $\mathbb{S}^2$, and the radiated field is sampled over angular observation directions on the sphere.

The spatial-sampling problem can be viewed as an inverse source problem, where a finite set of field samples over $\regR$ is used to estimate the source distribution over $\regT$ and reconstruct the complete observation field. The sampling performance is therefore determined by how well the selected measurements preserve the radiating source subspace~\cite{li2026near}. Let $\M H$ denote the densely sampled radiation operator, and $\M P_N$ denote the set of $N$ selected observation points. The corresponding sampled operator is denoted by
$\tM H=\M P_N \M H$~\cite{li2026near}.

Let the reduced singular value decomposition of the densely sampled radiation operator be
\(\M H=\M U\M\Sigma\M V^{\T H}\), where
\(\M\Sigma=\T{diag}(\sigma_1,\ldots,\sigma_{r_\T H})\),
\(r_\T H=\T{rank}(\M H)\), and the nonzero singular values are indexed in nonincreasing order. The worst-case reconstruction error over all unit-$L^2$ source distributions defines the spectral-norm sampling error~\cite{li2026near},
\begin{equation}
\mathcal{E}_{2}(\M P_N)=\|\M H(
\tM H^{+}\tM H-\M 1)\|_2\geq\sigma_{N+1},
\label{eq:spectral_sampling_error}
\end{equation}
where $\sigma_{N+1}$ is the $(N+1)$th singular value of $\M H$. Hence, the spectral norm characterizes the largest reconstruction error over all possible unit-norm source distributions.

To characterize the average reconstruction error over uncorrelated source-coefficient vectors satisfying \(\langle\M I_\T s\M I_{\T s}^{\T{H}}\rangle=\M{1}\), we  consider the Frobenius norm,
\begin{equation}
\mathcal{E}_{\T F}(\M P_N)=
    \langle \lVert\widehat{\M E}_{\T o}-\M E_{\T o}\rVert^2 \rangle_{\M I_\T s}^{1/2}
=\|\M H(\tM H^{+}\tM H-\M{1})\|_{\T F}.
\label{eq:frobenius_sampling_error}
\end{equation}
Using \(\|\M A\|_{\T F}^2=\T{tr}(\M A^{\T H}\M A)\)
and the cyclic property of the trace, the squared Frobenius error becomes
\begin{equation}
\mathcal{E}_{\T F}^{2}(\M P_N)=\T{tr}[\M H^{\T H}\M H(\M 1-\tM H^{+}\tM H)].
\label{eq:frobenius_trace_form}
\end{equation}

The Frobenius reconstruction error can be expressed in terms of the singular modes as
\begin{equation}
\mathcal{E}_{\T F}^{2}(\M P_N)=\sum_{n=1}^{r_\T H}\sigma_n^2[
1-\M v_n^{\T H}\tM H^{+}\tM H\M v_n]\geq \sum_{n=N+1}^{r_\T H}\sigma_n^2.
\label{eq:frobenius_modal_error}
\end{equation}
where $\V v_n$ denotes the $n$th right singular vector of $\M H$. \(\M v_n^{\T H}\tM H^{+}\tM H\M v_n\) measures the fraction of the $n$th source mode preserved by the sampled measurements. Since \(\tM H^{+}\tM H\)
projects onto a source subspace of dimension at most $N$, the lower bound follows from the Eckart--Young theorem~\cite{eckart1936approximation}.

The normalized Frobenius reconstruction error is
\begin{equation}
    \overline{\mathcal E}_{\T F}(\M P_N)=\frac{\|\M H(\tM H^{+}\tM H-\M 1)\|_\T{F}}{
    \|\M H\|_{\T F}}\geq ({{
\displaystyle\sum_{n>N}\sigma_n^2}/{\displaystyle\sum_n\sigma_n^2}})^{1/2}.
    \label{eq:far_field_fro_error}
\end{equation}
The corresponding normalized $L^2$-norm error is~\cite{li2026near}
\begin{equation}
    \overline{\mathcal E}(\M P_N)=\frac{\|\M H(\tM H^{+}\tM H-\M 1)\|_2}{\sigma_1}\geq\frac{\sigma_{N+1}}{\sigma_1}.
    \label{eq:far_field_2norm_error}
\end{equation}
The spectral and Frobenius norms therefore provide complementary measures: the former characterizes the worst-case error, while the latter captures the accumulated error over the singular modes. The Frobenius norm is particularly informative when the post-knee singular-value tail decays slowly.

\section{Far-field sampling}
\label{sec:far_field_sampling}

Far-field spatial sampling arises in applications such as antenna radiation-pattern measurements~\cite{antennas2021ieee}, over-the-air characterization~\cite{de2012lte}, radar and remote sensing~\cite{long2015microwave}, and direction-domain channel measurements~\cite{zwick2004novel}. In these scenarios, the observation distance is sufficiently large that the common radial dependence of the field can be separated from its angular dependence. The observation region can therefore be represented by directions on the unit sphere.

For far-field observation regions $\varOmega_\T o$ (in the far field, $\varOmega_\T o=\mathbb{S}^2$, see Fig.~\ref{fig:research_problem_surface}), the number of spatial DoFs, $\mathcal N_{\T {a}}$, characterizes the knee position separating the propagation modes from the reactive ones. Its dominant term is given by~\cite{gustafsson2025shadow}
\begin{equation}
\mathcal N_{\T {a}} = \frac{A_\T{os}}{\lambda^2},
\label{eq:Ndof}
\end{equation}
where \(\lambda\) is the wavelength and $A_\T{os}$ denotes the total shadow area associated with the source region $\regT$ over all observation directions,
\begin{equation}
A_\T{os} = \int_{\mathbb{S}^2} A_\T{s}(\UV{r}) \diff\Omega,
\label{eq:mutual_shadow}
\end{equation}
and $A_\T{s}(\UV{r})$ is the shadow (projected) area of $\regT$ in the direction $\UV{r}$. Consequently, the directional contribution to \(\mathcal N_{\T {a}}\), and hence the required sampling density, is proportional to
\({A_\T{s}(\UV{r})}/{\lambda^2}\) per polarization DoF. For symmetric configurations, both equivalent electric and magnetic currents are required to consistently account for the available DoF~\cite{gustafsson2025degrees}.

This sampling density provides a direct interpretation from an antenna perspective. For an idealized aperture with unit aperture efficiency, the directional gain is proportional to its projected area. The sampling density can therefore be interpreted as being proportional to the antenna gain \(G(\UV{r})\) in the corresponding direction,
\begin{equation}
\rho_\T{a}(\UV{r}) = \frac{A_\T{s}(\UV{r})}{\lambda^2}
\propto G(\UV{r}),
\end{equation}
up to an appropriate normalization factor. A larger projected source area generally corresponds to higher idealized gain and richer spatial bandwidth, and thus to a higher density of spatial DoFs. This establishes a direct connection to beamforming, since high-gain directions require denser sampling, whereas low-gain directions can be sampled more sparsely.

Hence, the sampling points should in general not be uniformly distributed with respect to the ordinary solid-angle measure $\diff\Omega$. Instead, they should be approximately uniformly distributed with respect to a weighted measure.

Based on the mutual-shadow measure~\eqref{eq:mutual_shadow}, the observation sphere is
partitioned into $N$ sampling cells with approximately equal DoF. For a prescribed number of samples, the target DoF associated with each cell is \(\Delta\mathcal{N}_{\T{a}}={\mathcal{N}_{\T{a}}}/{N}\). For $N$ sampling points, an ideal partition of the observation sphere into cells $\mathcal V_i$ satisfies
\begin{equation}
    \int_{\mathcal V_i}\frac{A_\T{s}(\UV{r})}{\lambda^2}    \diff\Omega\simeq\frac{\mathcal N_{\T{a}}}{N},\qquad i=1,\ldots,N.
    \label{eq:EqualShadowCells}
\end{equation}Equivalently, for a fixed geometry, the sampling density can be scaled by varying the wavelength, since \(\mathcal{N}_{\T{a}}\propto 1/\lambda^2\).

\begin{figure}[t]
  \centering
  \includegraphics[width=1\linewidth]{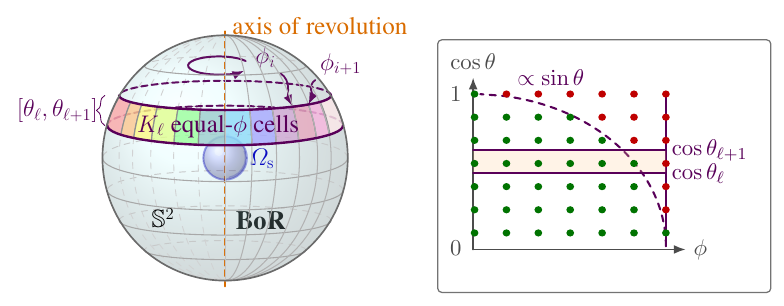}
\caption{Schematic of the far-field DoF sampling construction for a BoR source. Left: the observation sphere is partitioned into polar-angle layers, with each layer divided uniformly in $\phi$ owing to rotational symmetry. Right: the corresponding construction in the equal-area coordinates $(\phi,\cos\theta)$, where the spherical area element is transformed into a uniform measure and the sampling points are distributed uniformly. The dashed curve represents the spherical factor proportional to $\sin\theta$. The green dots denote the samples that map onto the spherical observation surface, where \(R_\ell\) samples are uniformly distributed over \(\phi\in[0,2\pi)\) on the \(\ell\)th latitude ring, while the red dots fall outside the spherical region.}
\label{fig:far_field_sampling_sphere}
\end{figure}

Body of revolution (BoR) geometries are widely encountered in practical antenna and scattering problems, including spherical, cylindrical, and other rotationally symmetric configurations~\cite{mautz1977h}. Owing to their rotational symmetry, they also provide a good starting point for illustrating the proposed DoF sampling construction. For a BoR, the directional DoF density is independent of the azimuthal angle and depends only on the polar angle, i.e., \(\rho_{\T{a}}=\rho_{\T{a}}(\theta)\). The sampling distribution can therefore be constructed layer by layer in the polar direction, with the samples within each layer uniformly distributed in azimuth.

As illustrated in Fig.~\ref{fig:far_field_sampling_sphere} (left), define the cumulative DoF from the north pole to the polar angle $\theta$ as
\begin{equation}
\mathcal N_{\T{a}}(\theta)=2\pi\int_{0}^{\theta}\rho_{\T{a}}(\theta)\sin\theta\,\diff\theta .
\label{eq:bor_cumulative_dof}
\end{equation}
Let the $l$th polar layer be bounded by $[\theta_l,\theta_{l+1}]$ and contain $R_l$ azimuthal cells. The DoF sampling method partitions the observation region into cells containing approximately one spatial DoF each, and one representative sampling point is placed in each cell. For a DoF sampling construction, one sampling point is first placed at each of the two poles. The remaining samples are then arranged in latitude layers, whose boundaries are selected such that
\begin{equation}
\mathcal N_{\T{a}}(\theta_{l+1})-\mathcal N_{\T{a}}(\theta_l)=R_l .
\label{eq:bor_layer_dof}
\end{equation}
Since $\rho_{\T{a}}$ is independent of $\phi$, the layer is then divided into $R_l$ equal azimuthal sectors with
\(\Delta\phi_l={2\pi}/{R_l}\). A sampling direction is then placed at the DoF-weighted center of each cell, for example
\begin{equation}
\theta_l^{\T s}=\mathcal N_{\T{a}}^{-1}({\mathcal N_{\T{a}}(\theta_l)+\mathcal N_{\T{a}}(\theta_{l+1})})/{2}
\label{eq:theta}
\end{equation}
and the azimuthal center
\begin{equation}
\phi_{l,m}^{\T s}=(m+{1}/{2})\Delta\phi_l,
\qquad m=0,\ldots,R_l-1.
\label{eq:phi}
\end{equation}
The \(\Delta\phi_l/2\)-cell offset is introduced to align each sampling point with the center of its corresponding cell.

The right panel of Fig.~\ref{fig:far_field_sampling_sphere} provides an equivalent representation in the $(\phi,\cos\theta)$ coordinates. Each polar layer appears as a horizontal strip bounded by $\cos\theta_l$ and $\cos\theta_{l+1}$. The $l$th layer contains $R_l$ spatial DoFs and is therefore assigned $R_l$ sampling points, which are uniformly distributed along $\phi$. This representation makes the variation of the azimuthal sampling density with polar angle explicit and illustrates how the DoF sampling cells remain approximately compact over the sphere.

We first consider a spherical source of radius $a=2\lambda$. For an electrically large spherical region, the number of spatial DoFs is approximately $(ka)^2$ per polarization~\cite{bucci1997electromagnetic}, where $k=2\pi/\lambda$ is the free-space wavenumber. Hence, $(ka)^2=(4\pi)^2\simeq158$, in agreement with the mutual-shadow estimate $\mathcal N_{\T{a}}\simeq158$ obtained from~\eqref{eq:Ndof}, which is used as the characteristic DoF knee in the reconstruction analysis. The scalar free-space Green's function $G(\V r,\V r^\prime)={\T{exp}(-\ju k|\V R|)}/({4\pi |\V R|})$ is used to construct the densely sampled channel matrix $\M H$ and its sampled channel $\tM H$.

\begin{figure}[t]
  \centering
  \includegraphics[width=0.8\linewidth]{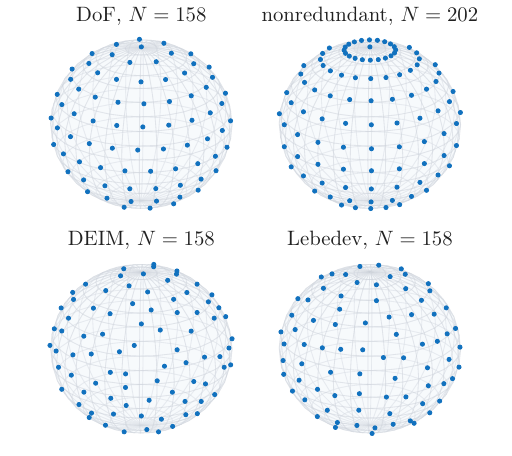}
\caption{
Sampling distributions for the spherical source with radius \(2\lambda\). The panels compare the proposed DoF sampling method, nonredundant sampling~\cite{bucci1998representation}, DEIM~\cite{hochman2014reduced}, and the Lebedev-based scheme~\cite{lebedev1999quadrature}.}
  \label{fig:far_field_samlping_distribution}
\end{figure}

Owing to the rotational symmetry of the spherical source, the shadow-area density is constant over the unit sphere $\mathbb{S}^2$. Accordingly, as shown in Fig.~\ref{fig:far_field_samlping_distribution},
the proposed DoF sampling construction yields a nearly uniform distribution with respect to the spherical-area measure. The resulting spatial distribution is next compared with the nonredundant~\cite{bucci1998representation}, the discrete empirical interpolation method (DEIM)~\cite{hochman2014reduced}, and Lebedev~\cite{lebedev1999quadrature} sampling schemes.

In the nonredundant/reduced sampling method, the asymptotic required number of samples $N_{\T{rs}}$ is determined by the area of the closed surface $\varOmega_\T s$ enclosing the source~\cite{bucci1998representation,Bucci2025,bucci1988use}
\begin{equation}
    N_{\T{rs}}=\frac{|\varOmega_\T s|}{(\lambda/2)^2}.
    \label{eq:bucci_count}
\end{equation}
The corresponding sampling grid is constructed directly in the angular coordinates. From~\eqref{eq:Ndof} and~\eqref{eq:bucci_count},
\(N_{\T{rs}}/\mathcal N_{\T{a}}\approx4/\pi\). For a prescribed target count, the numbers of polar and azimuthal samples are chosen such that the actual number of points is close to the target while the angular steps \(\Delta\theta\) and \(\Delta\phi\) remain comparable. The samples are then placed on uniformly spaced polar rings with uniformly spaced azimuthal angles, as illustrated in Fig.~\ref{fig:far_field_sampling_sphere} and Fig.~\ref{fig:far_field_samlping_distribution}. Although this construction is simple and geometry independent, it is uniform in the angular coordinates \((\theta,\phi)\) rather than in physical area on the sphere. Since \(\diff\Omega=\sin\theta\,\diff\theta\diff\phi\), equal angular increments correspond to progressively smaller solid-angle elements toward the poles. Consequently, the sampling density per unit solid angle increases in the polar regions, which is evident from both the sampling geometry in Fig.~\ref{fig:far_field_sampling_sphere} and the corresponding point distribution in Fig.~\ref{fig:far_field_samlping_distribution}, leading to substantial sampling redundancy near the poles.

 DEIM~\cite{hochman2014reduced} is used as a matrix-based benchmark. DEIM follows a nested greedy selection procedure, in which the addition of a new sampling point does not alter the locations of the previously selected points. Its nested structure allows the sampling set to be progressively enriched, making it convenient for evaluating reconstruction performance for different~\(N\). For a spherical source, Lebedev quadrature~\cite{lebedev1999quadrature} can also be directly employed.

In Fig.~\ref{fig:far_field_samlping_distribution}, DEIM uses the same number of displayed samples as the DoF sampling scheme and provides comparable global coverage, although its greedy algebraic selection leads to a less regular spatial pattern. Finally, the Lebedev scheme also produces an approximately uniform spherical-area distribution and exhibits similar global coverage. However, its sampling points are distributed in an irregular, non-trajectory-based manner, whereas the proposed DoF sampling scheme organizes the samples systematically along latitude layers, making it better suited to practical spherical scanning procedures~\cite{hansen1988spherical}.

\begin{figure}[t]
  \centering
  \includegraphics[width=1\linewidth]{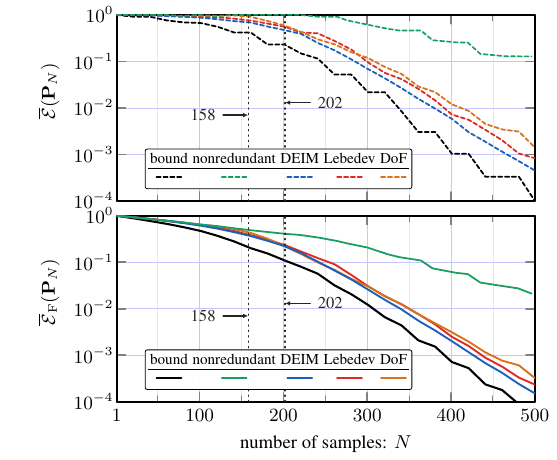}
\caption{Normalized spectral- and Frobenius-norm reconstruction errors versus the number of far-field samples $N$ for a spherical source of radius $2\lambda$, comparing DoF sampling, nonredundant, DEIM, and Lebedev sampling schemes with the lower bounds in~\eqref{eq:far_field_fro_error} and~\eqref{eq:far_field_2norm_error}. The vertical dotted lines mark the estimated $\lceil\mathcal N_{\T{a}}\rceil=158$ from~\eqref{eq:Ndof}, and the nonredundant sampling~\cite{bucci1998representation}, $N\approx202$.}
  \label{fig:spectra_far_field}
\end{figure}
As shown in Fig.~\ref{fig:spectra_far_field}, the DoF sampling, Lebedev, and DEIM schemes exhibit similar convergence trends and closely follow the reconstruction lower bounds in~\eqref{eq:far_field_fro_error}, particularly under the Frobenius norm. This agreement is notable because the proposed DoF sampling points are determined solely from the geometry-dependent shadow measure, without requiring an operator-dependent point-selection stage. It therefore motivates a geometry-based construction in which each sampling cell contains approximately one spatial DoF and is represented by a single sample. The resulting performance indicates that the geometry-based construction captures the relevant far-field spatial mode using only geometrical information. Compared with the spectral norm, the Frobenius norm is better suited to this case because the singular-value spectrum exhibits a relatively slowly decaying post-knee tail, for which the cumulative contribution of multiple residual modes is more representative than the largest residual mode alone. Accordingly, the Frobenius norm is adopted for the reconstruction-error evaluation in the remainder of this paper.

The nonredundant sampling scheme converges more slowly. Around its nominal critical count,
$N_{\T{rs}}\simeq202$, both reconstruction errors remain significantly larger than those of the DoF sampling and DEIM schemes. This is consistent with the point clustering seen in Fig.~\ref{fig:far_field_samlping_distribution}, samples are over-allocated near the poles and under-allocated around the equatorial region in terms of physical area. The comparison therefore emphasizes that the spatial allocation of the samples is as important as the nominal sample count.

To examine whether the normalized sampling behavior is specific to a spherical source, we next consider three families of convex regions: cylinders, spheres, and prolate ellipsoids. Within each family, all source dimensions are uniformly scaled, while preserving the shape and aspect ratio to realize $\mathcal N_{\T{a}}=10$, $50$, and $200$. Specifically, from~\eqref{eq:Ndof}, the wavelength is determined as \(\lambda=\sqrt{{A_\T{os}}/{\mathcal N_{\T{a}}}}\). This allows the sampling behavior to be compared consistently across different source shapes and electrical sizes without altering their physical geometry.

\begin{figure}[t]
  \centering
  \includegraphics[width=1\linewidth]{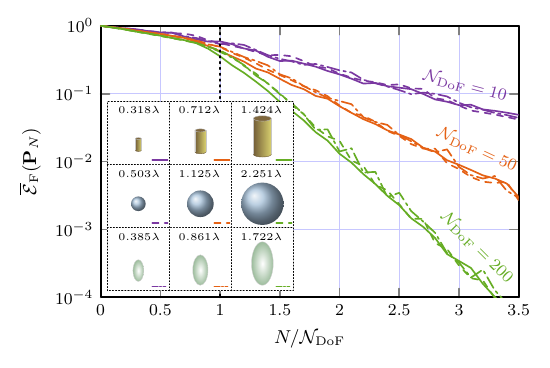}
\caption{Normalized Frobenius reconstruction error for DoF far-field sampling of closed cylindrical, spherical, and prolate-ellipsoidal source surfaces. In the inset, the top row shows cylinders with height $4a$, where $a$ denotes the cylinder radius; the middle row shows spheres characterized by their radius $a$; and the bottom row shows prolate ellipsoids with equal transverse semi-axes $a$ and polar semi-axis $2a$. The numerical labels indicate the corresponding values of $a$ for each source geometry.}
  \label{fig:spectra_far_field_cylinder}
\end{figure}
As shown in Fig.~\ref{fig:spectra_far_field_cylinder}, the normalized reconstruction error decreases consistently with increasing $N/\mathcal N_{\T{a}}$ for all considered source geometries. More importantly, the decay becomes substantially faster as $\mathcal N_\T a$ increases. For larger \(\mathcal N_{\T{a}}\), the reconstruction error exhibits a pronounced knee as \(N/\mathcal N_{\T{a}}\) increases, followed by a continued decrease over several orders of magnitude under further oversampling. In contrast, the lower-DoF cases exhibit a noticeably slower decay over the same normalized sampling range. This indicates that the asymptotic DoF sampling construction becomes increasingly accurate for electrically larger sources.

For a fixed $\mathcal N_{\T{a}}$, the error curves associated with the different source geometries remain close over most of the sampling range. This confirms that $\mathcal N_{\T{a}}$ captures the dominant scaling of the required number of samples, despite substantial differences in source shape. The curves do not, however, collapse exactly onto a universal profile. The remaining deviations arise from the geometry-dependent distribution of the mutual-shadow density and the projected aperture widths over the observation sphere, which determine the local sampling-cell area and sampling-cell aspect ratio, respectively. These variations are more pronounced for non-spherical geometries, particularly the prolate ellipsoid, leading to greater sensitivity of the discrete layer assignment and consequently to small fluctuations in the reconstruction-error curve.

\begin{figure}[t]
  \centering
\includegraphics[width=0.95\linewidth]{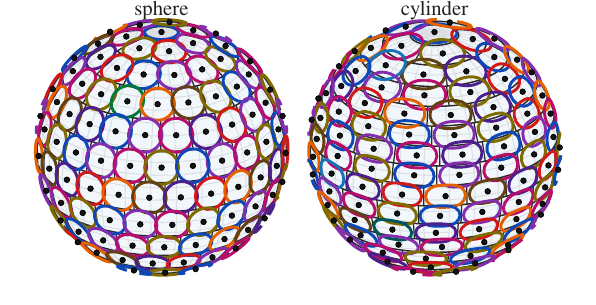}
\caption{Beam regions over the observation sphere for the proposed DoF sampling. The grid delineates the DoF sampling partition of the observation sphere, with each cell representing approximately one spatial DoF and the black dot indicating its associated sampling direction. It shows the MRT beam half-amplitude contours for all sampling directions. The source geometries are a sphere of radius $2.251\lambda$ and a cylinder of radius $1.424\lambda$ and height $5.696\lambda$ (see inset in Fig.~\ref{fig:spectra_far_field_cylinder}). The NDoF $\mathcal{N}_{\T{a}}=200$ with $\lambda$ from~\eqref{eq:Ndof} is used.}
\label{fig:far_field_beam}
\end{figure}

Fig.~\ref{fig:far_field_beam} illustrates the beam regions associated with the proposed DoF sampling directions for the sphere and the cylinder, considered in Fig.~\ref{fig:spectra_far_field_cylinder}. For each sampling direction, maximum-ratio transmission (MRT)~\cite{maisto2021near} produces a focused beam that can equivalently be interpreted as a spatial point-spread function (PSF) centered at the corresponding observation direction~\cite{rao2001performance}. Accordingly, each sampling direction defines a beam region, and the collection of these beam regions provides approximately continuous coverage of the observation sphere. For the spherical source, the nearly isotropic projected aperture produces regular and approximately isotropic beam regions, whereas the cylindrical source gives rise to direction-dependent and elongated beam regions due to its anisotropic projected aperture. The resulting stretching ratio between the $\theta$- and $\phi$-directions is consistent with the local cell aspect ratio estimated from the DoF sampling method.

\section{Near-field surface sampling}\label{sec:Near-field surface sampling}

Near-field spatial sampling is important in applications such as antenna measurements~\cite{yaghjian1986overview}, electromagnetic imaging~\cite{broquetas1998spherical}, and large-aperture communication systems~\cite{dardari2020communicating}. In this regime, the field variation depends strongly on the source--observation geometry, motivating a geometry-adaptive sampling strategy. As in the far-field case, the spatial DoF is governed by the geometrical projection between the source and observation regions. In the far field, this projection depends only on the observation direction, whereas in the near field it also varies with the relative position and orientation of the source and observation surface elements. Based on the mutual-shadow framework~\cite{gustafsson2025shadow,brick2026interpreting}, which relates the total DoF to the mutual-shadow area, we introduce a DoF density through the incremental mutual-shadow contribution associated with an infinitesimal observation-surface element. 

Assuming that each source point $\V r^\prime$ is visible from every observation point $\V r$, let $\V R=\V r-\V r^\prime$ and define the  propagation direction
$\UV R=\V R/|\V R|$. The mutual shadow area between the source and
observation surfaces is
\begin{equation}
A_{\T{os}}=\int_{\regR}\int_{\regT}\frac{
|\UV R\cdot\UV n|\,|\UV R\cdot\UV n^\prime
|}{|\V R|^2}\,\diff \T S^\prime \diff \T S ,
\label{eq:mutual_shadow_area}
\end{equation}
where $\UV n^\prime$ and $\UV n$ are the unit normals of the source
and observation surfaces, respectively. The corresponding asymptotic spatial DoF is
$\mathcal N_{\T{a}}=A_{\T{os}}/\lambda^2$ as in~\eqref{eq:Ndof}.

By expressing the total DoF as an integral over the observation surface,
\(\mathcal N_{\T{a}}=\int_{\regR}\rho_{\T{a}}\,\diff \T S\), the mutual-shadow DoF density follows as
\begin{equation}
\rho_{\T {a}}=\frac{1}{\lambda^2}\int_{\regT}\frac{|\UV R\cdot\UV n|\,
|\UV R\cdot\UV n^\prime|}{|\V R|^2}\,\diff \T S^\prime .
\label{eq:dof_density_integral_gradient}
\end{equation}
The DoF density determines the area of the local sampling cell. For a fixed physical
geometry, its normalized spatial distribution is frequency independent, whereas the area of a cell scales as $\lambda^2$.

For compact and well-separated source--observation configurations and with the observation-surface normal $\UV n$ chosen to
point toward the source region, the propagation direction
$\UV R$ from the source to the observation point satisfies \(\UV R\cdot\UV n < 0\) for all relevant source--observation pairs. The sign of this projection is therefore fixed, such that \(|\UV R\cdot\UV n|=-\UV R\cdot\UV n\). Under this condition, the absolute value associated with the observation normal in~\eqref{eq:dof_density_integral_gradient} can be removed, and the DoF density can be written as a linear form \begin{equation}
\rho_{\T {a}}= \frac{-\UV n \cdot}{\lambda^2}\int_{\regT}\frac{
|\UV R\cdot\UV n^\prime|\,\UV R}{|\V R|^2}\,\diff \T S^\prime=\UV n\cdot\V s(\V r),
\label{eq:dof_density}
\end{equation}
where the contributions from the entire transmitting aperture are collected into the DoF density vector
\begin{equation}
\V s(\V r)=-\frac{1}{\lambda^2}\int_{\regT}\frac{|\UV R\cdot\UV n^\prime|\,\UV R}{|\V R|^2}\,\diff \T S^\prime.
\label{eq:effective_propagation_vector}
\end{equation}
The fixed-sign condition is essential for this representation. If
$\UV R\cdot\UV n$ changes sign over the interacting source--observation
pairs, the absolute value in~\eqref{eq:dof_density_integral_gradient}
cannot in general be removed, and the DoF density cannot be represented
by the above DoF density vector. The associated DoF density direction is defined as $\UV s=\V s/|\V s|$, which provides a convenient local representation of the combined propagation contributions from the transmitting aperture. In the remainder of this paper, we consider configurations in which \(\V R\cdot\UV n\) has a consistent sign over the mutually visible source--observation pairs, as assumed in \eqref{eq:dof_density}. Such special cases require a separate treatment and are therefore beyond the scope of this work. Their characterization and
corresponding sampling strategies are left for future study.

\begin{figure}[t]
  \centering
\includegraphics[width=1\linewidth]{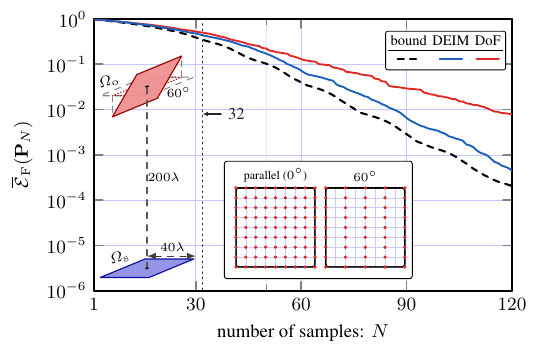}
\caption{Normalized Frobenius reconstruction error for the $60^\circ$-tilted Rx geometry. The insets show the source--observation geometry, where the source and observation apertures are both $40\lambda\times 40\lambda$, together with the analytical sampling grids for the parallel and $60^\circ$-tilted receive apertures.}
\label{fig:paraxial regime}
\end{figure}

In the far field, the observation surface is perpendicular to the propagation direction. For near-field sampling over a general surface, however, the DoF density direction varies across the observation surface and need not be normal to it. To separate the effects of propagation direction from those of sampling density, we first consider a paraxial configuration for which the DoF density is approximately constant. In this regime, the DoF density direction $\UV s$ is approximately constant over the observation aperture and plays a role analogous to that of the radial direction $\UV r$ in far-field sampling.

Consider two parallel, well-separated planar apertures, as shown in Fig.~\ref{fig:paraxial regime}. The DoF density direction is approximately parallel to the aperture normals, $\UV s\approx\pm\UV n$, 
and hence its tangential projection vanishes. The sampling can therefore be realized by an approximately uniform grid along any two orthogonal tangential directions, e.g., $\UV t_1=\UV x$ and $\UV t_2=\UV y$ for separation along the $\UV z$ direction.

The DoF density \(\rho_{\T{a}}^{(0)}\) and NDoF \(\mathcal N_{\T{a}}^{(0)}\) are approximately~\cite{miller2019waves}
\begin{equation}
\rho_{\T{a}}^{(0)} \simeq \frac{|\regT|}{\lambda^2d^2}
\quad\text{and }
\mathcal N_{\T{a}}^{(0)} \simeq\frac{|\regT||\regR|}{\lambda^2d^2},
\end{equation}
where $d$ is the center-to-center separation. For
$|\regT|=|\regR|=(40\lambda)^2$ and $d=200\lambda$, this gives
$\mathcal N_{\T{a}}^{(0)}\simeq64$, as illustrated in Fig.~\ref{fig:paraxial regime}.

\begin{figure}[t]
  \centering
\includegraphics[width=1\linewidth]{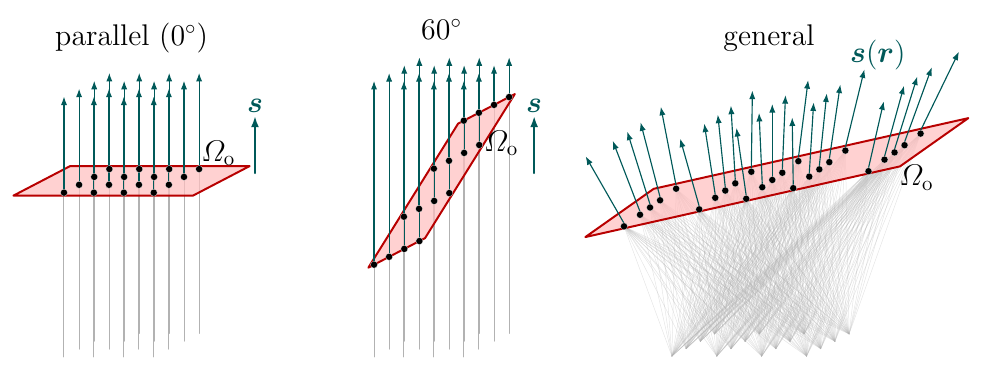}
\caption{Field-line interpretation between the DoF density direction \(\V s(\V r)\) and the observation region \(\regR\). From left to right: the parallel and $60^\circ$-tilted aperture configurations considered in Fig.~\ref{fig:paraxial regime}, and a general source--observation geometry. The arrows represent the field lines associated with the DoF density direction $\V s(\V r)$, and the black markers denote their intersections with $\regR$, interpreted as sampling locations. The gray lines below $\regR$ illustrate the contributions from the transmitting region to these directions.}
\label{fig:vector_field_surface}
\end{figure}

More generally, the vector field $\V{s}(\V r)$ provides a geometric
interpretation of the proposed sampling construction. Outside the
transmitting region $\regT$, $\V{s}(\V r)$ is solenoidal, with field lines originating from $\regT$. Their intersections with the observation region $\regR$ can be interpreted as the corresponding sampling locations, as illustrated in
Fig.~\ref{fig:vector_field_surface} for the parallel and tilted paraxial configurations of Fig.~\ref{fig:paraxial regime}, together with a general source--observation geometry. This viewpoint is closely related to the nonredundant sampling framework~\cite{bucci1998representation,Bucci2025,bucci1988use}
and extends to the present 2D observation surfaces, the local density of field-line intersections is governed by the DoF density, while their orientation relative to the observation surface is determined by the DoF density direction of $\V{s}(\V r)$.

Next, rotate the observation aperture by an angle $\theta$ about the tangential direction $\UV t_1$. The DoF density direction $\UV s$ remains approximately constant over $\regR$, but is no longer normal to the observation aperture. As illustrated in the middle panel of Fig.~\ref{fig:vector_field_surface}, the intersections of the field lines with $\regR$ consequently become anisotropically distributed. The sampling spacing remains unchanged along the rotation axis $\UV t_1$, whereas it is stretched along the orthogonal tangential direction $\UV t_2=\UV n\times\UV t_1$.

The stretching factor is $1/|\cos\theta|$. For example, at $\theta=60^\circ$, the sampling spacing along $\UV t_2$ doubles. Equivalently, the projected observation area is reduced by a factor $\cos\theta$, so that the NDoF decreases from approximately $64$ to $32$. This behavior is illustrated in Fig.~\ref{fig:paraxial regime}.

This paraxial example suggests a local sampling construction for a general source--observation geometry. We define two orthogonal in-plane sampling directions,
\begin{equation}
\UV t_1 ={\UV s\times\UV n}/{|\UV s\times\UV n|}
\quad\text{and }
\UV t_2 =\UV n\times\UV t_1,
\end{equation}
where $\UV t_2$ is aligned with the tangential projection of $\UV s$, while $\UV t_1$ is orthogonal to it. The corresponding projection factors are
\begin{equation}
    p_1=1
    \quad\text{and }
    p_2=|\UV s\cdot\UV n|.    
\end{equation}

Let $h_1$ and $h_2$ denote the local dimensions of a sampling cell along $\UV t_1$ and $\UV t_2$, respectively. Requiring equal effective projected sampling resolution in the two directions gives
\begin{equation}
    {h_1}/{h_2} =|\UV s\cdot\UV n|.
\end{equation}
The absolute dimensions $h_1$ and $h_2$ are then determined by the
mutual-shadow DoF density in~\eqref{eq:dof_density_integral_gradient},
such that each sampling cell represents approximately one spatial DoF. Thus, the DoF density determines the local cell area, while the projection factor $|\UV s\cdot\UV n|$ determines its aspect ratio and the directions $\UV t_1$ and $\UV t_2$ determine its orientation. In a general geometry, $\V{s}(\V r)$ varies over the observation region, so that these quantities vary locally across $\regR$. This is illustrated in the right panel of
Fig.~\ref{fig:vector_field_surface}, where the spatially varying field lines intersect $\regR$ at nonuniformly distributed locations.

In the limiting case $|\UV s\times\UV n|\rightarrow0$, the propagation direction is normal to the surface and no preferred in-plane direction exists. The sampling cells are then locally isotropic, and any orthonormal tangential basis may be used.

The complete sampling pattern is generated by successively constructing neighboring cells using the locally updated directions $\UV t_1$ and $\UV t_2$. The construction starts from regions with high \(\rho_{\T {a}}\) and progressively expands toward regions with lower \(\rho_{\T {a}}\), since the high-density regions contribute more strongly to the dominant singular-value content of the channel and should therefore be resolved first. When the DoF density direction is locally normal to the receiving surface, the corresponding sampling location is first determined by integrating one DoF. The remaining sampling cells are then constructed in regions where the tangential projection of $\UV s$ is nonzero.

\begin{figure}[t]
  \centering
\includegraphics[width=1\linewidth]{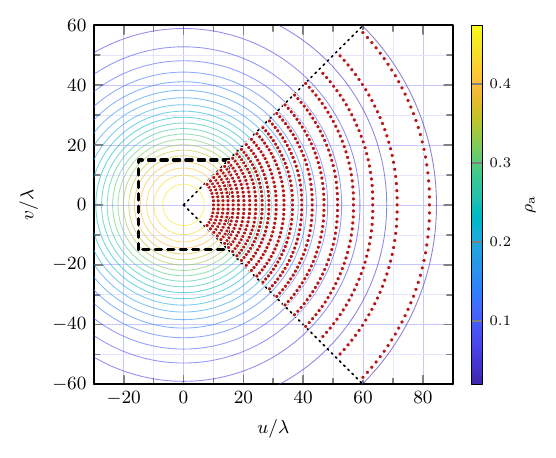}
\caption{Geometry-adaptive DoF spatial sampling over the local DoF-density distribution. The background contours represent the local DoF density $\rho_{\T{a}}$. Starting from the prescribed central point $(10\lambda,0\lambda)$, corresponding to the central angular position, each sampling layer is traced in both clockwise and counterclockwise directions along the corresponding iso-DoF-density contour. The associated cell size determines the local sampling interval, and the tracing proceeds symmetrically toward the two angular limits at $\pm45^\circ$. The dashed square denotes the $30\lambda\times30\lambda$ transmitting aperture.}
\label{fig:point_density_distribution}
\end{figure}
As shown in Fig.~\ref{fig:point_density_distribution}, the sampling points exhibit a clear spatially varying density that follows the nonuniform DoF distribution. In the region closer to the transmitting aperture, where the local DoF density is higher and varies more rapidly, the samples are packed more closely. As the distance from the aperture increases, the DoF density gradually decreases and the spacing between adjacent sampling layers becomes larger. A further observation is that the sampling trajectories progressively spread out while preserving an ordered structure across the different layers. Consequently, fewer samples are required in regions with lower spatial variation, whereas denser sampling is retained only where it is necessary.

\begin{figure}[t]
  \centering
\includegraphics[width=0.8\linewidth]{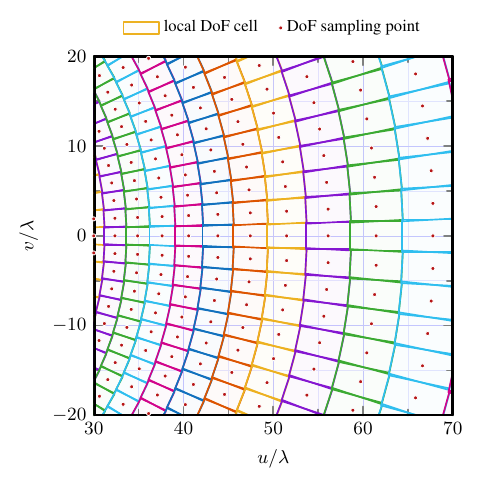}
\caption{Local DoF sampling partition of the observation region from Fig.~\ref{fig:point_density_distribution} based on the DoF sampling, where the displayed subregion uses the same coordinate range as that selected in Fig.~\ref{fig:point_density_distribution}.}
\label{fig:rectangular_1dof_partition}
\end{figure}
The partition in Fig.~\ref{fig:rectangular_1dof_partition} further confirms that the nonuniform sampling pattern arises directly from the spatial variation of the DoF structure. In particular, the gradual change in the effective region associated with neighboring samples is consistent with the variation in sampling spacing observed in Fig.~\ref{fig:point_density_distribution}. This correspondence intuitively explains why the proposed construction can redistribute a fixed number of samples according to the sampling density, allocating fewer samples to regions of low DoF density and more to regions of high DoF density. Since the partition is obtained numerically over a finite spatial grid, each cell may exhibit a small deviation from the ideal one-DoF value. However, these deviations are minor and have a negligible influence on the resulting sampling distribution.

\begin{figure}[t]
  \centering
\includegraphics[width=1\linewidth]{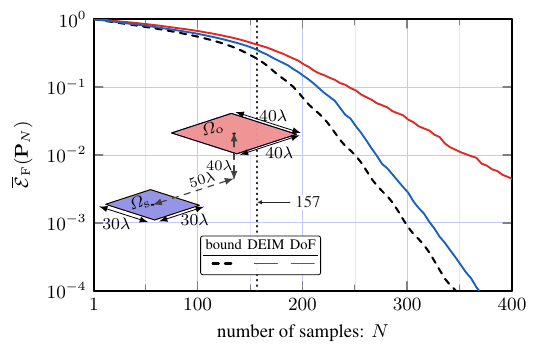}
\caption{Normalized Frobenius reconstruction error for offset square apertures. The proposed DoF sampling is compared with DEIM and the lower bound~\eqref{eq:far_field_fro_error}, with $\lceil\mathcal N_{\T{a}}\rceil=157$ calculated from ~\eqref{eq:Ndof}.}
\label{fig:singular_value_spectra_surface}
\end{figure}
Fig.~\ref{fig:singular_value_spectra_surface} compares the normalized Frobenius reconstruction error of the proposed DoF sampling with DEIM and the corresponding lower bound~\eqref{eq:far_field_fro_error} for the offset square apertures. The estimate gives $\lceil\mathcal N_{\T{a}}\rceil=157$ from~\eqref{eq:Ndof}, which approximately marks the onset of the faster error-decay region. Before this point, all three curves decrease relatively slowly, whereas beyond the DoF estimate the lower bound and DEIM exhibit a much steeper decay. The proposed DoF sampling follows the same overall trend but with a more gradual decay, resulting in a higher reconstruction error than DEIM at larger sample numbers. Nevertheless, the error decreases consistently as additional samples are introduced, showing that the sampling distribution captures the dominant spatial information. The remaining gap to the lower bound is mainly attributed to edge effects associated with the finite source and observation regions~\cite{li2026near}, which are not explicitly accounted for in the present DoF construction.

\begin{figure}[t]
  \centering
\includegraphics[width=1.03\linewidth]{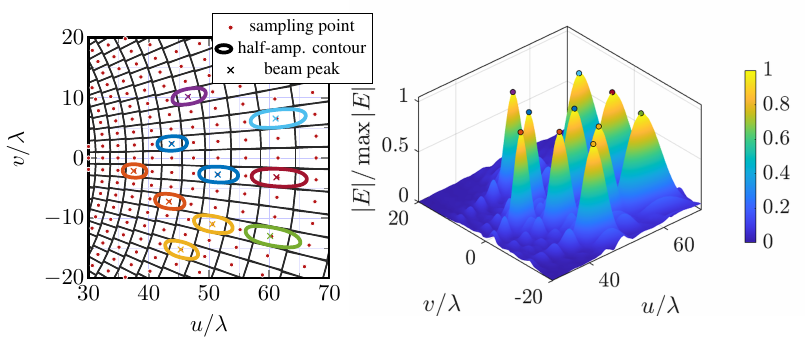}
\caption{MRT beam regions for a set of selected focal points. The left panel shows the sampling points, beam peaks, and each beam's half-amplitude contour on the observation plane, while the right panel shows the corresponding 3D normalized field distribution. The setup is the same as that shown in the inset of Fig.~\ref{fig:singular_value_spectra_surface}.}
\label{fig:beam_zf_verification}
\end{figure}
Large-aperture communication systems require spatial channel information over an extended surface for beamforming, channel characterization, and spatial processing. To further assess whether the proposed DoF sampling preserves the spatial characteristics required for beamforming, Fig.~\ref{fig:beam_zf_verification} shows the beam regions obtained using MRT for a set of selected focal points. The MRT beams form compact half-amplitude beam regions centered around their corresponding focal points, indicating that the sampled channel retains the local focusing capability of the continuous aperture. The neighboring beam regions remain well localized with limited spatial spreading, demonstrating that distinct focal locations can be resolved over the observation surface. The beam stretching ratio along the two local
tangent directions is also consistent with the ratio $h_1/h_2$. The corresponding 3D normalized field distribution is shown in the right panel of Fig.~\ref{fig:beam_zf_verification}. Each MRT beam is normalized to its own maximum. The peaks are clearly separated over the observation surface, while the field magnitude decays rapidly away from the focal regions. This confirms that the individual MRT beams are spatially localized around their respective target positions, with limited overlap between their dominant field regions.

\begin{figure}[t]
  \centering
  \includegraphics[width=1\linewidth]
{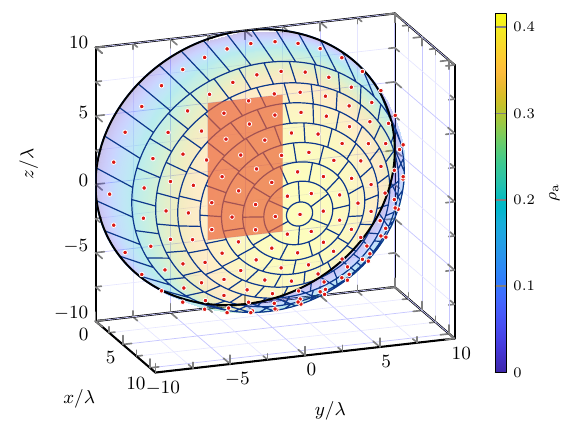}
\caption{DoF-density distribution and the proposed DoF sampling over a hemispherical observation surface. The planar source aperture is $5\lambda\times10\lambda$, and the hemisphere has radius $10\lambda$. The color map represents $\rho_{\T{a}}$, while the blue curves and red dots denote the integrated cell boundaries and sampling locations, respectively. The corresponding DoF estimate is
$\lceil\mathcal{N}_{\T{a}}\rceil=158$, as obtained from~\eqref{eq:Ndof}.}
\label{fig:spherical_sampling}
\end{figure}

As a representative near-field spherical measurement application~\cite{hansen1988spherical}, we consider a hemispherical observation surface surrounding a planar transmitting aperture. The resulting DoF-density distribution and the proposed integrated DoF sampling pattern are shown in Fig.~\ref{fig:spherical_sampling}. The DoF density is strongly geometry dependent, with larger values in the region directly facing the source aperture and a gradual decrease toward the outer part of the hemisphere. Accordingly, the observation surface is partitioned into approximately equal-DoF cells, resulting in smaller cells in high-density regions and progressively larger cells in low-density regions. To obtain a more regular sampling pattern, the cells are organized in successive layers and adjacent layers are staggered in the angular direction, avoiding persistent radial alignment of the cell boundaries and sampling points. This construction is appropriate for the considered geometry, but it is not necessarily directly applicable to arbitrary source--observation configurations. For the considered $5\lambda\times10\lambda$ planar source and
hemisphere of radius $10\lambda$, the mutual-shadow measure gives $\lceil\mathcal N_{\T{a}}\rceil = \lceil\pi|\regT|/\lambda^2\rceil=158$, which is used as the DoF sampling count.

\begin{figure}[t]
  \centering
  \includegraphics[width=1\linewidth]
{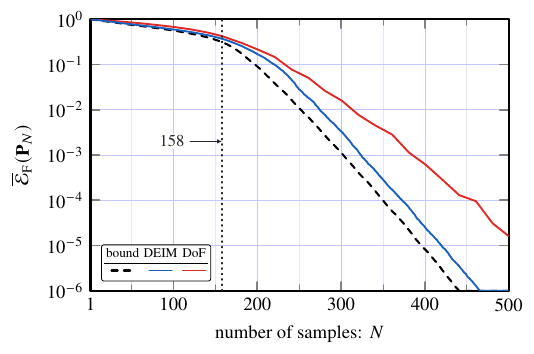}
\caption{Normalized Frobenius reconstruction error versus the number of samples for the hemispherical near-field observation geometry. The proposed DoF sampling is compared with DEIM and the lower bound~\eqref{eq:far_field_fro_error}. The vertical dotted line indicates the DoF-based sample count $\lceil\mathcal N_{\T{a}}\rceil=158$.}
\label{fig:spherical_sampling_spectra}
\end{figure}
Fig.~\ref{fig:spherical_sampling_spectra} compares the normalized Frobenius reconstruction errors of the proposed DoF sampling and DEIM for the hemispherical observation geometry. Beyond the DoF knee, the proposed DoF sampling distribution continues to provide stable convergence. Incorporating an appropriate edge correction into the DoF sampling construction is expected to further reduce the reconstruction error, particularly in the post-knee region~\cite{li2026near}.

The proposed method relies on the leading-order mutual-shadow DoF~\eqref{eq:Ndof} and is therefore most suitable for electrically large systems. Finite-size and boundary effects are neglected. In addition, the local anisotropy is represented by a single DoF density direction, so strongly multidirectional fields may require a higher-order angular description.

\section{Conclusions}\label{sec:Conclusions}
This paper developed a geometry-based spatial sampling framework for far- and near-field electromagnetic propagation using the mutual-shadow DoF measure. In the far field, the source shadow area determines the sampling density over the observation sphere, leading to an approximately equal-DoF partition of the angular domain. In the near field, the mutual-shadow DoF density determines the local sampling-cell area, while the DoF density direction determines the cell orientation and aspect ratio, resulting in a geometry-adaptive sampling pattern over the observation surface. Numerical results show that the estimated DoF identifies the characteristic transition in the reconstruction error and that the resulting DoF sampling distributions preserve the dominant spatial information. The framework therefore provides a unified geometry-based sampling strategy for both far- and near-field configurations. Future work may focus on developing a more general edge-correction strategy to account for the finite boundaries of both the source and observation regions, thereby further improving the reconstruction accuracy for sampling.

\bibliographystyle{IEEEtran}
\bibliography{citations}

\vfill

\end{document}